\documentclass[aps,floatfix,showpacs,superscriptaddress,showkeys,preprint]{revtex4}
\usepackage{graphicx,epsfig,epstopdf}
\usepackage{amssymb,amsmath,amsxtra,amsfonts}
\usepackage{bm}
\usepackage{color}

\usepackage{titlesec}
\titleformat{\section}{\large\bfseries}{\thesection}{1em}{}

\newcommand{\bea}{\begin{eqnarray}}
\newcommand{\ena}{\end{eqnarray}}
\newcommand{\be}{\begin{equation}}
\newcommand{\en}{\end{equation}}

\newcommand{\ed}{\end{document}}

\begin{document}

\hfill MITP/16-083 (Mainz)                

\title{Isospin-violating strong decays of scalar single-heavy tetraquarks} 

\author{Thomas Gutsche}
\affiliation{
Institut f\"ur Theoretische Physik, Universit\"at T\"ubingen,
Kepler Center for Astro and Particle Physics, 
Auf der Morgenstelle 14, D-72076, T\"ubingen, Germany}

\author{Mikhail A. Ivanov}
\affiliation{Bogoliubov Laboratory of Theoretical Physics,
Joint Institute for Nuclear Research, 141980 Dubna, Russia}

\author{J\"{u}rgen G. K\"{o}rner}
\affiliation{PRISMA Cluster of Excellence, Institut f\"{u}r Physik, 
Johannes Gutenberg-Universit\"{a}t, 
D-55099 Mainz, Germany}

\author{Valery E. Lyubovitskij}
\affiliation{
Institut f\"ur Theoretische Physik, Universit\"at T\"ubingen,
Kepler Center for Astro and Particle Physics, 
Auf der Morgenstelle 14, D-72076, T\"ubingen, Germany}
\affiliation{ 
Department of Physics, Tomsk State University,  
634050 Tomsk, Russia} 
\affiliation{Laboratory of Particle Physics, Mathematical Physics Department, 
Tomsk Polytechnic University, Lenin Avenue 30, 634050 Tomsk, Russia} 

\begin{abstract}

We present a study of the isospin-violating one-pion strong decays of 
single heavy tetraquarks $X_Q = X(s q \bar q \bar Q)$, $Q=c, b,$ and $q=u, d$ 
with spin-parity $J^P = 0^+$. We assume  that the 
tetraquarks have the configuration of a color diquark and 
an antidiquark. Three mechanisms of isospin
violation can contribute to the decay rate:
(1) mixing of the $X(s u \bar u \bar Q)$ and $X(s d \bar d \bar Q)$ 
tetraquark currents,
(2) an explicit $m_d-m_u$ quark mass difference in the quark diagrams 
describing the corresponding decay transitions, and  
(3) $\pi^0-\eta$ mixing in the final state. 
Our main results are as follows: 
(1) It is quite likely that the investigated tetraquark states are 
isosinglet states with a small admixture of an isotriplet component; 
(2) The first isospin-breaking mechanism affects
the decay rather more significantly than the others; 
(3) Our calculations contain a size parameter $\Lambda_{X_Q}$, characterizing 
the distribution of the quarks in the tetraquark state $X_Q$. 
Absolute decay rates depend very much on the choice of $\Lambda_{X_Q}$, 
varied from 1 to 2 GeV, reflecting the compactness of the multiquark system. 
 
\end{abstract}


\maketitle

\section{Introduction}

The past years have been marked by extensive experimental and 
theoretical studies of the $X$, $Y$, and $Z$ states or
of heavy mesons containing at least one heavy $c$ or $b$ quark.
Many of these reported resonances cannot easily be explained as
quark-antiquark configurations; alternative structure interpretations
involve, for example, hadron molecules or compact tetraquark states
among others.

In the present study we focus on the scalar resonance $D_{s0}^\ast(2317)$
and its possible partner state $B_{s0}^\ast$. 
We use a tetraquark picture to analyze the
isospin-violating one-pion strong decays of 
the $D_{s0}^\ast(2317)$  and its bottom companion $B_{s0}^\ast$. 
Assuming tetraquark configurations these states contain a single heavy 
quark ($c$ or $b$), a strange quark and a pair of nonstrange quarks. 

The $D_{s0}^\ast(2317)$ and $B_{s0}^\ast$ states have been studied before 
in detail using different theoretical approaches, including the hadronic 
molecular approach (see overview and references 
in~\cite{Faessler:2007gv}-\cite{Agashe:2014kda}), where these 
states are considered to be bound states of the $D$ and $K$, $B$ and $K$,
respectively.
In most of the approaches, the $D_{s0}^\ast(2317)$ and $B_{s0}^\ast$ states 
have been considered to be isosinglet states, which is consistent 
with recent results by the Belle Collaboration~\cite{Choi:2015lpc}.  
The Belle Collaboration reported that 
possible isotriplet partners of the $D_{s0}^{\ast}(2317)$ have not been found. 
In Ref.~\cite{Faessler:2007gv} it was proposed 
that in the framework of the hadronic molecular picture 
there are two mechanisms for the pion emission in the reaction 
$D_{s0}^\ast \to D_s + \pi^0$: 
(1)~a~direct mechanism due to the emission from the $(DK)$ loop; 
(2) the $\eta-\pi^0$ mixing transition. 
It was shown that the direct transition
dominates over the $\eta-\pi^0$ mixing transition in the
$D_{s0}^{\ast} \to D_s \pi^0$ decay. In Ref.~\cite{Faessler:2008vc} 
the formalism proposed in Ref.~\cite{Faessler:2007gv} has been 
extended to the case of the $B_{s0}^\ast$ (bound state of $B$ and $K$ mesons) 
--- bottom partner of the $D_{s0}^{\ast}$ state. In calculations performed in 
Refs.~\cite{Faessler:2008vc,Faessler:2007gv} the distribution of 
hadronic constituents in $D_{s0}^\ast$ and $B_{s0}^\ast$ was described 
by a scale parameter $\Lambda$, which was found to be of order of 1 GeV. 
In particular, in Refs.~\cite{Faessler:2007gv}-\cite{Faessler:2008vc}
it was calculated that 
$\Gamma(D_{s0}^{\ast} \to D_s + \pi^0) = 46.7 - 111.9$ keV and
$\Gamma(B_{s0}^{\ast} \to B_s + \pi^0) = 55.2 - 89.9$ keV
when the scale parameter describing the distribution of constituents
in the hadronic molecule was varied from 1 to 2 GeV. 

In the present manuscript we return to 
the problem of the isospin-violating decays 
$D_{s0}^{\ast} \to D_s \pi^0$ and $B_{s0}^{\ast} \to B_s \pi^0$. 
A new feature in our study is that we apply the covariant confined 
multiquark approach proposed and developed 
in Refs.~\cite{Dubnicka:2010kz}. This method is an 
extension of the covariant relativistic quark 
model~\cite{Ivanov:1996pz} devised for 
a unified description of bound state structures of hadrons 
and exotic states. As in the hadronic molecular approach one has 
a free scale parameter $\Lambda$, which is fixed from the 
description of the decay rates 
$\Gamma(D_{s0}^{\ast} \to D_s + \pi^0)$ and 
$\Gamma(B_{s0}^{\ast} \to B_s + \pi^0)$.  
We find that our $\Lambda \sim 1$ GeV is compatible with the 
scale parameter found in the framework of the hadonic molecular
picture~\cite{Faessler:2007gv}-\cite{Faessler:2008vc}.

\section{Formalism}

Our starting point is that $D_{s0}^{\ast}$ and $B_{s0}^{\ast}$ states are 
bound states of four quarks having the configuration of a color diquark and 
an antidiquark. In our phenomenological Lagrangian formalism such a
configuration is 
encoded in the tetraquark interpolating currents discussed 
in detail in the literature (see e.g. 
recent review~\cite{Nielsen:2009uh}) and, in particular, 
in the context of the covariant tetraquark 
confinement model~\cite{Dubnicka:2010kz}. 

For our specific cases of 
the $D_{s0}^{\ast +} = (c q) (\bar q \bar s)$ and the 
    $B_{s0}^{\ast 0} = (s q) (\bar q \bar b)$ tetraquark states 
we construct the currents in the form of the mixed 
isosinglet $J^S$ and isotriplet $J^T$ (third component) 
currents 
\bea\label{current_mixing} 
J_{D_{s0}^{\ast +}} &=& \cos\delta \, J_{D_{s0}^{\ast +}}^S 
                     +  \sin\delta \, J_{D_{s0}^{\ast +}}^T 
\,,\nonumber\\
J_{B_{s0}^{\ast 0}} &=& \cos\delta \, J_{B_{s0}^{\ast 0}}^S 
                     +  \sin\delta \, J_{B_{s0}^{\ast 0}}^T 
\,.
\ena 
The singlet-octet mixing angle $\delta$ parametrizes the isospin 
violation in the structure of the $D_{s0}^{\ast}$ or $ B_{s0}^{\ast}$ states. 
We keep $\delta$ as a free parameter. 
The isosinglet $J^S$ and isotriplet $J^T$ currents are defined as 
\bea 
J_{D_{s0}^{\ast +}}^{S/T} &=& 
\frac{1}{\sqrt{2}} \Big[ J(c u \bar u \bar s) \pm J(c d \bar d \bar s) \Big]
\,, \\
J_{B_{s0}^{\ast 0}}^{S/T} &=& 
\frac{1}{\sqrt{2}} \Big[ J(s u \bar u \bar b) \pm J(s d \bar d \bar b) \Big]
\,. 
\ena 
The color structure of a generic tetraquark current 
$J(q_1 q_2 \bar q_3 \bar q_4)$ has 
the configuration of a color diquark-antidiquark with 
\bea 
J(q_1 q_2 \bar q_3 \bar q_4) &=& D^c_{12} \, D^{c \dagger}_{34} = 
\varepsilon^{abc} \varepsilon^{dec} \, 
\bigg[ q_1^a C \Gamma_1 q_2^b \biggr]  \, 
\bigg[ \bar q_3^d \Gamma_2 C \bar q_4^e \biggr] \,.  
\ena 
where 
\bea 
D^c_{12} = \varepsilon^{abc} \, 
\bigg[ q_1^a C \Gamma_1 q_2^b \biggr]  \,.  
\ena 
The indices $a,b,c,d,e$ refer to color, $C = \gamma^0 \gamma^2$ 
is the charge conjugation matrix, and $\Gamma_1$ and $\Gamma_2$ are 
the Dirac spin matrices resulting in zero total angular momentum and positive 
$P$ parity for the interpolating tetraquark current. In particular, 
the following combinations of the ($\Gamma_1,\Gamma_2$) matrices 
(without involving derivatives) are possible:  
\bea 
P &=& \Gamma_1 \otimes \Gamma_2 = \gamma^5 \otimes \gamma^5 \,, 
\nonumber\\ 
S &=& \Gamma_1 \otimes \Gamma_2 = I \otimes I\,, 
\nonumber\\ 
A &=& \Gamma_1 \otimes \Gamma_2 = \gamma^5\gamma^\mu \otimes 
\gamma_\mu\gamma^5 \,, 
\nonumber\\ 
V &=& \Gamma_1 \otimes \Gamma_2 = \gamma^\mu \otimes \gamma_\mu \,, 
\nonumber\\ 
T &=& \Gamma_1 \otimes \Gamma_2 = 
\frac{1}{2} \sigma^{\mu\nu} \gamma_5 \otimes 
\sigma_{\mu\nu} \gamma_5 \,.
\ena 
When we take the heavy quark limit for the $b$ constituent, which is
equivalent to 
the nonrelativistic limit, the $S$ current vanishes, 
the $P$ and $A$ currents are degenerate resulting in 
the spin structure $\sigma^2 \otimes \sigma^2$, and the 
$V$ and $T$ currents are also degenerate producing the spin 
structure $\sigma^2 \sigma^i \otimes \sigma^i \sigma^2$. 
In this paper, for simplicity, we work with the simplest $P$ current.  
From our experience based on
analysis of single heavy baryons (see Ref.~\cite{Ivanov:1999pz})  
the observables are not so sensitive to a choice of the interpolating current.
Therefore, we do not expect that the use of $V(T)$ or mixing of two possible
currents $P(A)$ and $V(T)$ for scalar tetraquarks with an extra free parameter
should drastically change the description of physical properties
of these exotic states. However, such an analysis could be done
in our future study. 
After having specified the color, spin and flavor structure of our 
tetraquark currents we are in the position to implement the coordinate 
(or space-time) part and construct phenomenological Lagrangians describing 
the interaction of the tetraquark states
$H = D_{s0}^{\ast +}, B_{s0}^{\ast 0}$
with their constituents. 
We proceed in complete analogy to the original work on the $X(3872)$ 
treated as a tetraquark state (see details in Refs.~\cite{Dubnicka:2010kz}). 
The interaction Lagrangian of the tetraquark states 
$H = D_{s0}^{\ast}, B_{s0}^{\ast}$ 
with their constituents is constructed as 
\bea  
{\cal L}_{H}(x) = g_{H} \, H(x) \, 
J_{H}(x) + {\rm H.c.} 
\ena  
$J_{H}(x)$ is the nonlocal confined tetraquark current 
including the appropriate spin, flavor and color structure discussed before. 
For example, a generic current $J_H$ for $H \equiv X(s q \bar q \bar Q)$ 
corresponding to the coupling of the  color diquark 
$D_1^c = \varepsilon^{abc} \, \bigg[ q^a C\gamma^5 s^b \bigg]$ 
and the antidiquark 
$D_2^{c \dagger} = 
\varepsilon^{dec} \, \bigg[ \bar q^d \gamma^5 C \bar Q^e \bigg]$ 
has the form 
\bea 
J_{X(s q \bar q \bar Q)}(x) &=& \int d^4x_1 \cdots \int d^4x_4 \, 
\delta\biggl( x - \sum\limits_{i=1}^4 w_i x_i \biggr) \nonumber\\ 
&\times&  
\Phi\biggl( \sum\limits_{i < j} (x_i - x_j)^2 \biggr) 
\varepsilon^{abc} \varepsilon^{dec} \ 
\bigg[ q_a(x_1) C \gamma^5 s_b(x_2) \biggr] \, 
\bigg[ \bar q_d(x_3) \gamma^5 C \bar Q_e(x_4) \biggr] \,, 
\ena 
where $\Phi$ is the correlation function of the $X(s q \bar q \bar Q)$ state 
providing for the ultraviolet finiteness of all matrix elements. 

The coupling constant $g_{H}$ is determined through 
the compositeness condition 
$Z_{H} = 1 - g_H^2 \Pi^\prime(M_H^2) = 0$, where 
$\Pi^\prime$ is the derivative of the tetraquark mass operator 
(the relevant diagram for the mass operator is displayed in Fig.1).  
The compositeness condition sets the 
wave function renormalization constant to zero, which means that 
the tetraquark state is a dressed bound state of four valence quarks.  

\begin{figure}[htb]
\begin{center}
\hspace*{-2.5cm}
\epsfig{figure=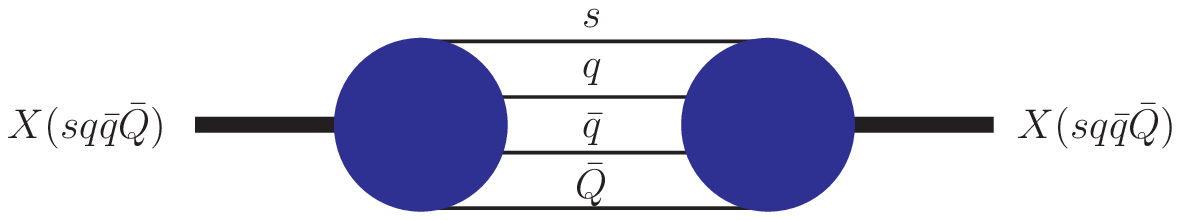,scale=.65}

\vspace*{-.25cm}
\caption{Mass operator of the tetraquark $X(s q  \bar q \bar Q)$  
\label{fig:mass_operator}}
\vspace*{.5cm}
\hspace*{-2.5cm}
\epsfig{figure=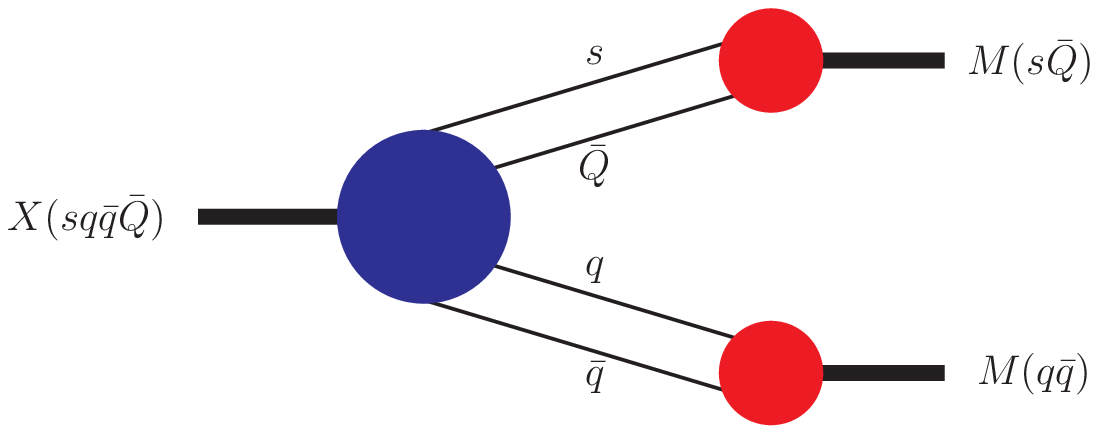,scale=.65}
\caption{Strong decay of the tetraquark $X(s q  \bar q \bar Q)$ into 
two mesons $M(s \bar Q)$ and $M(q \bar q)$ 
\label{fig:decay}}
\end{center}
\end{figure}

The explicit expression for the derivative of the mass operator
$\Pi^\prime(p^2)$ 
corresponding to Fig.~1 reads 
\bea 
\Pi^\prime(p^2) &=& \frac{1}{2p^2} \, p^\alpha \, 
\frac{\partial}{\partial p^\alpha} \Pi(p^2) = 
\frac{4N_c}{2p^2} \, F(m_Q,m_s,m_q) \,,
\ena 
where $4N_c = 12$ is the appropriate color factor and 
the structure integral  $F(m_Q,m_s,m_q)$ is given by 
\bea 
F &=& 
\prod\limits_{i=1}^3 \int \frac{d^4k_i}{(2 \pi)^4 i} 
\, \Phi^2(-K^2) \nonumber\\
&\times& \biggl[ 
- w_s 
{\rm tr}[S_s^{[12]} \not\! p S_s^{[12]} \gamma^5 S_{q}^{[2]} \gamma^5] 
\, 
{\rm tr}[S_Q^{[3]} \gamma^5 S_{q}^{[13]} \gamma^5] \nonumber\\
&-& w_Q 
{\rm tr}[S_s^{[13]} \gamma^5 S_{q}^{[2]} \gamma^5] \, 
{\rm tr}[S_Q^{[3]} \not\! p S_Q^{[3]} \gamma^5 S_{q}^{[13]} \gamma^5] 
\nonumber\\
&+& w_{q} 
{\rm tr}[S_s^{[12]} \gamma^5 S_{q}^{[2]} \not\! p S_{q}^{[2]} \gamma^5] \, 
{\rm tr}[S_Q^{[3]}  \gamma^5 S_{q}^{[13]} \gamma^5] 
\nonumber\\
&+& w_{q} 
{\rm tr}[S_s^{[12]} \gamma^5 S_{q}^{[2]} \gamma^5] \, 
{\rm tr}[S_Q^{[3]}  \gamma^5 S_{q}^{[13]} \not\! p S_{q}^{[13]} \gamma^5] 
\biggr]\,. 
\ena 
The free constituent quark propagators 
$1/(m_q - \not\! k)$ are denoted by  $S_q^{[\ldots]}$
with the specific momentum dependence 
\bea 
& & S_s^{[12]} = S_s(k_1 + k_2 - w_s p)\,, \quad 
    S_Q^{[3]}  = S_Q(k_3 - w_Q p)\,, \nonumber\\ 
& & S_{q}^{[2]}  = S_{q}(k_2 + w_u p)\,, \quad 
    S_{q}^{[13]} = S_{q}(k_1 + k_3 + w_d p)\,, 
\ena 
where $w_i = m_i/(2 m_{q} + m_s + m_Q)$ is the fractional quark mass. 
The distribution of the constituent quarks in the bound state
is modeled by the correlation 
function $\Phi(-K^2)=\exp(K^2/\Lambda^2)$ with the momentum dependence 
\bea 
K^2  = \frac{1}{8} (k_1 + 2k_2)^2 
     + \frac{1}{8} (k_1 + 2k_3)^2 
     + \frac{1}{4} k_1^2 
\ena 
where  $\Lambda$ is a free parameter. 

We are now in the position to discuss the calculation of the 
isospin-violating one-pion strong decay of the 
tetraquark states summarized by the Feynman diagram of Fig.~2.
As partially emphasized in the Introduction and at the beginning of 
this section there are several mechanisms that contribute 
to the amplitude of this isospin-violating decay process.
The first mechanism (I) refers to the isospin violation in the structure
of the  $D_{s0}^{\ast}$ and $B_{s0}^{\ast}$ states. It is already parametrized
by the singlet-octet mixing angle present in the tetraquark currents of 
Eq.~(\ref{current_mixing}).
The second mechanism (II) includes isospin breaking corrections based on
the $u$ and $d$ quark mass difference. We work with the $u$ and $d$ 
quark masses,  
\bea 
\tilde m_u = m - \frac{\Delta}{2}\,, \ 
\tilde m_d = m + \frac{\Delta}{2}\,, 
\ena 
where $m= m_u = m_d$ is their value in the isospin limit and 
$\Delta = \tilde m_d - \tilde m_u$ is the $d-u$ quark mass difference. 
For an estimate of $\Delta$ we take a value  
given by the difference $M_d - M_u$ 
of the $d$ and $u$ current quark masses as listed 
by the PDG~\cite{Agashe:2014kda},  
\bea\label{Delta_def} 
\Delta = \tilde m_d - \tilde m_u = M_d - M_u = 2.5 \ {\rm MeV}\,, 
\ena  
To implement the third mechanism (III) we take into account the $\pi^0-\eta$ 
mixing in the interaction Lagrangians of these mesons with 
their constituent quarks. As shown in Ref.~\cite{Gasser:1984gg} 
the $\pi^0$ and $\eta$ meson fields are modified by a unitary transformation
given by 
\bea\label{Unit_trans}
\pi^0 \ \to \ \pi^0 \cos\varepsilon - \eta \sin\varepsilon 
\,,  \quad 
\eta  \ \to \ \pi^0 \sin\varepsilon + \eta \cos\varepsilon
\ena
where $\varepsilon$ is the $\pi^0-\eta$ mixing angle fixed
by~\cite{Gasser:1984gg}:
\bea\label{tan2e}
\tan 2\varepsilon = \frac{\sqrt{3}}{2}\, 
\frac{M_d-M_u}{M_s - \frac{M_u+M_d}{2}} 
\simeq 0.024  
\ena
and where $M_q$ are the current quark masses. 
For the above estimate of $\tan 2\varepsilon$ 
we use the central values for the current 
quark masses from the PDG~\cite{Agashe:2014kda}: 
\bea 
   M_d - M_u = 2.5 \ {\rm MeV}\,, \ 
   M_u + M_d = 7.6 \ {\rm MeV}\,, \ 
   M_s = 95 \ {\rm MeV}\,.
\ena  
As a result of the unitary transformation~(\ref{Unit_trans})
the neutral pion couples to both isosinglet 
$\bar u u + \bar d d$ and isotriplet 
$\bar u u - \bar d d$ currents 
\bea 
{\cal L}_{\pi}(x) = \frac{1}{\sqrt{2}} \, 
\pi^0(x) \int d^4y \, \Phi_\pi(y^2) \, 
\biggl[ g_\pi 
\cos\varepsilon J_{q\bar q}^-(x,y) +  
\frac{g_\eta}{\sqrt{3}} 
\sin\varepsilon J_{q\bar q}^+(x,y) \biggr] 
\ena 
where 
\bea 
J_{q\bar q}^\pm(x,y) &=&
\bar u(x+y/2) i\gamma^5 u(x-y/2) \nonumber\\
&\pm& 
\bar d(x+y/2) i\gamma^5 d(x-y/2) \,.
\ena  

Combining all three isospin breaking contributions we obtain 
the following expression for the amplitude of strong  
$D_{s0}^{\ast \pm} \to D_s^\pm + \pi^0$ 
and $B_{s0}^{\ast 0}(\bar B_{s0}^{\ast 0}) 
\to B_s^{0}(\bar B_s^0) + \pi^0$ decays:  
\bea\label{delta_eps}  
{\cal B} &=& \frac{{\cal B}_{u\bar u}}{2} \Big( 1 + 
\sin\delta + 
\frac{g_{\eta}}{g_{\pi} \sqrt{3}} \sin\epsilon \Big)  \nonumber\\
&+&\frac{{\cal B}_{d\bar d}}{2} \Big( - 1 + \sin\delta 
+ \frac{g_{\eta}}{g_{\pi} \sqrt{3}} \sin\epsilon \Big)  \nonumber\\
&=& \frac{{\cal B}_{u\bar u} - {\cal B}_{d\bar d}}{2} 
+ \Big(\sin\delta + \sin\varepsilon\frac{g_{\eta}}{g_{\pi} \sqrt{3}} \Big)  
\, \frac{{\cal B}_{u\bar u} + {\cal B}_{d\bar d}}{2} \,. 
\ena 
The quantity 
${\cal B}_{q\bar q}$ is the amplitude of the one-pion strong isospin-violating 
decay of the tetraquark with two specific light nonstrange quarks $u \bar u$ 
or $d \bar d$,  
\bea
{\cal B}_{q\bar q} 
= 6 g_{X_{s q \bar q \bar Q}} g_{M(s \bar Q)} g_\pi 
G(m_Q,m_s,m_q,m_q) \,. 
\ena 
The last equation corresponds to the amplitude of the transition 
$X(s q \bar q \bar Q) \to M(s \bar Q) + M(q \bar q)$ and 
is defined as the product of a color factor $N_c! = 6$, 
constants describing the respective couplings of the tetraquark 
$X(s q \bar q \bar Q)$, meson $M(s \bar Q)$, and pion $g_\pi$ 
with the constituent quarks and the mass-dependent structure integral 
$G(m_Q,m_s,m_q)$. 
The latter is given by   
\bea 
G &=& 
\int \frac{d^4k_1}{(2 \pi)^4 i} \, 
\int \frac{d^4k_2}{(2 \pi)^4 i} \, 
\Phi(-L^2) \, \Phi_1(-L_1^2) \, 
\Phi_2(-L_2^2) \nonumber\\
&\times& 
{\rm tr}[\gamma^5 S_s(k_1) \gamma^5 S_Q(k_1+q_1) 
\gamma^5 S_u(k_2) \gamma^5 S_d(k_2+q_2)] \,, 
\nonumber 
\ena 
where 
\bea 
L^2 &=& \frac{1}{8} [ 2 k_1 + q_1 (1 + w_Q - w_s)   
+ q_2 (w_Q-w_s) ]^2 \nonumber\\
&+& \frac{1}{8} [ 2 k_2 + q_2 ]^2 
+ \frac{1}{4}[ 2 q_1 w_q - q_2 (w_Q - w_s]^2 \,, \nonumber\\
L_1^2 &=& [k_1 + \tilde w_Q]^2 \,, \ \ 
L_2^2 \ = \ [k_2 + 1/2]^2 
\ena 
and $\tilde w_Q = m_Q/(m_Q+m_s)$. 
Here $\Phi_1(-L_1^2) = \exp(L_1^2/\Lambda_{M_1}^2)$ and 
$\Phi_2(-L_2^2) = \exp(L_2^2/\Lambda_{M_2}^2)$ are 
the correlation functions of the mesons $M(s\bar Q)$ and 
$M(q\bar q)$, respectively. 

The strong decay width of the scalar tetraquark state $H$ into 
two pseudoscalar mesons $M_1$ and $M_2$ is 
evaluated according to the expression 
\bea 
\Gamma(H \to M_1 + M_2) = 
\frac{\lambda^{1/2}(M_H^2,M_1^2,M_2^2)}{16 \pi M_X^3} 
\, |{\cal B}|^2 \,. 
\ena

Most of the model parameters have been fixed in 
previous calculations: the constituent quark masses 
$m_u=m_d=241$ MeV, $m_s=428$ MeV, $m_c=1.672$ GeV, 
$m_b=5.046$~GeV, the scale parameters of the interaction vertex with 
$\Lambda_\pi = 0.871$~GeV, 
$\Lambda_\eta = 1$~GeV,
$\Lambda_{D_s} = 1.81$~GeV,
$\Lambda_{B_s} = 2.05$~GeV, and the infrared 
confinement scale parameter $\lambda = 0.181$~GeV. 
The scale parameters of the tetraquark states are free parameters. 
The coupling constants of the pion $g_{\pi}$, $\eta$ meson $g_\eta$, 
and $D_s$ and $B_s$ mesons $g_{D_s}$ and $g_{B_s}$ 
have also been evaluated in previous calculations 
(see, e.g., Ref.~\cite{Dubnicka:2010kz}): 
$g_{\pi} = 5.18\,, g_{\eta} = 4.15\,, 
g_{D_s} = 3.76\,, g_{B_s} = 4.97\,.$ 
Finally, we remind reader that three scenarios are specified 
by the following choice of isospin-breaking parameters [see definitions 
in Eqs.~(\ref{current_mixing}), (\ref{Delta_def}), 
(\ref{Unit_trans}), and (\ref{tan2e})] and strong amplitude of 
one-pion transition ${\cal B}$: 
  
Scenario I 
\bea
& &\sin\delta = 0.012 \,, \ \ \Delta \equiv 0 \,, \ \ 
\sin\varepsilon \equiv 0 \,, 
\nonumber\\ 
& &{\cal B} = \sin\delta 
\, \frac{{\cal B}_{u\bar u} + {\cal B}_{d\bar d}}{2} \,, 
\ena 
Scenario II 
\bea
& &\sin\delta \equiv 0\,, \ \ \Delta = 2.5 \, \mathrm{MeV} \,, \ \
\sin\varepsilon \equiv 0 \,, \nonumber\\
& &{\cal B} = 
\frac{{\cal B}_{u\bar u} - {\cal B}_{d\bar d}}{2} \,, 
\ena 
Scenario III 
\bea
& &\sin\delta \equiv 0\,, \ \ \Delta \equiv 0  \,, \ \
 \sin\varepsilon = 0.012 \,, \nonumber\\
& &{\cal B} = \sin\varepsilon \frac{g_{\eta}}{g_{\pi} \sqrt{3}} 
\, \frac{{\cal B}_{u\bar u} + {\cal B}_{d\bar d}}{2} \,, 
\ena 
Full result including all mechanisms of isospin breaking I+II+III 
\bea
& &\sin\delta = \sin\varepsilon = 0.012\,, \ \ \Delta = 2.5 \, \mathrm{MeV} 
\,, \nonumber\\
& &{\cal B} = \frac{{\cal B}_{u\bar u} - {\cal B}_{d\bar d}}{2} 
+ \Big(\sin\delta + \sin\varepsilon\frac{g_{\eta}}{g_{\pi} \sqrt{3}} \Big)  
\, \frac{{\cal B}_{u\bar u} + {\cal B}_{d\bar d}}{2} \,. 
\ena 
One can see that in the numerical analysis 
we use approximation 
$\sin\delta \simeq \sin\varepsilon = 0.012$.

\section{Results} 

Our numerical results for the isospin-violating decay rates of the scalar
single-heavy tetraquarks are shown in Tables I - IV. 
We present our results for different scenarios; i.e.,    
we first restrict ourselves to a specific isospin breaking mechanism
I, II, or III and then calculate  the full result, including all
mechanisms I+II+III. 
From the results  
it is obvious that the mixing of the isosinglet and isotriplet tetraquark 
currents has the largest effect on the decay rate rather than 
the pure $d-u$ quark mass difference or the $\pi^0-\eta$ mixing. 

We furthermore vary the scale parameter $\Lambda$ within 
a reasonable range of values 
from 1 to 2~GeV. This follows estimates done for the scale parameter 
$\Lambda_{X_b}\sim 1.4 -2$~GeV of the hypothesized single-heavy 
tetraquark $X(5568)$ as performed in Ref.~\cite{Goerke:2016hxf}. 
A larger value of $\Lambda$ would correspond to a compact tetraquark 
configuration, and a $\Lambda$ close to 1 GeV is also reflected in 
configurations with a larger extent like hadronic molecules.
In the case of the $B_{s0}^{\ast 0}$ state its mass is 
varied from 5.725 to 6.3 GeV, and for the $D_{s0}^{\ast +}$ the mass 
is fixed at 2.317 GeV.
With $\Lambda \approx 1$ GeV the one-pion decay rates of the isosinglet 
tetraquark states with a small admixture of the isotriplet 
tetraquark component are actually compatible with results 
in the hadonic molecular 
picture~\cite{Faessler:2007gv}-\cite{Faessler:2008vc} 
giving predictions for the isospin-breaking 
decay rate of the order of 100 keV. 
In particular, in Refs.~\cite{Faessler:2007gv}-\cite{Faessler:2008vc} 
we found that 
$\Gamma(D_{s0}^{\ast} \to D_s + \pi^0) = 46.7 - 111.9$~keV and 
$\Gamma(B_{s0}^{\ast} \to B_s + \pi^0) = 55.2 - 89.9$~keV 
when the scale parameter describing the distribution of constituents 
in the hadronic molecule is varied from 1 to 2~GeV. 

As we mentioned before, we present numerical 
results using approximation 
$\sin\delta \simeq \sin\varepsilon$ or 
$R = \frac{\sin\delta}{\sin\varepsilon} \simeq 1$. 
For completeness we also give our predictions for arbitrary values  
of the ratio parameter $R$ for two limiting values of the scale  
parameter $\Lambda_H = 1$ and $2$ GeV, respectively, 
\bea 
& &\Gamma(D_{s0}^{\ast} \to D_s + \pi^0) = 6.17 \, (1 + 1.82 \, R)^2 \ 
\mathrm{keV}\,, \nonumber\\
& &\Gamma(B_{s0}^{\ast} \to B_s + \pi^0) = 5.15 \, (1 + 1.83 \, R)^2 \  
\mathrm{keV} \ \mathrm{at} \ M_{B_{s0}^{\ast 0}} = 5.725 \  
\mathrm{GeV} \,, \nonumber\\
& &\Gamma(B_{s0}^{\ast} \to B_s + \pi^0) = 6.36 \, (1 + 1.77 \, R)^2 \  
\mathrm{keV} \ \mathrm{at} \ M_{B_{s0}^{\ast 0}} = 6.1 \  
\mathrm{GeV} \,, \nonumber\\
& &\Gamma(B_{s0}^{\ast} \to B_s + \pi^0) = 5.49 \, (1 + 1.73 \, R)^2 \  
\mathrm{keV} \ \mathrm{at} \ M_{B_{s0}^{\ast 0}} = 6.3 \ 
\mathrm{GeV} 
\ena 
and 
\bea 
& &\Gamma(D_{s0}^{\ast} \to D_s + \pi^0) = 0.46 \, (1 + 1.83 \, R)^2 \ 
\mathrm{keV}\,, \nonumber\\
& &\Gamma(B_{s0}^{\ast} \to B_s + \pi^0) = 0.36 \, (1 + 1.84 \, R)^2 \  
\mathrm{keV} \ \mathrm{at} \ M_{B_{s0}^{\ast 0}} = 5.725 \  
\mathrm{GeV} \,, \nonumber\\
& &\Gamma(B_{s0}^{\ast} \to B_s + \pi^0) = 0.63 \, (1 + 1.87 \, R)^2 \  
\mathrm{keV} \ \mathrm{at} \ M_{B_{s0}^{\ast 0}} = 6.1 \  
\mathrm{GeV} \,, \nonumber\\
& &\Gamma(B_{s0}^{\ast} \to B_s + \pi^0) = 0.71 \, (1 + 1.88 \, R)^2 \  
\mathrm{keV} \ \mathrm{at} \ M_{B_{s0}^{\ast 0}} = 6.3 \ 
\mathrm{GeV} \,.
\ena 
The present calculation gives predictions for the one-pion decay rates 
of about $3 - 5$ keV when the configuration is chosen to be more compact.
Absolute rates clearly depend on the size of the tetraquark configuration, 
because of the lack of data especially for the $D_{s0}^{\ast}(2317)$ 
the two scenarios, a rather compact or an
extended configuration, cannot be distinguished.

\section{Summary and conclusions}

Let us summarize the main results of our paper. 
We have considered the isospin-violating decays of the
isosinglet states with a small admixture of the isotriplet component.
We found that for values of the scale parameter of the order of 1 GeV 
the one-pion decay rates of these tetraquark states are compatible 
with results found within the hadronic molecular approach. On the other 
hand, an increase in the scale parameter up to 2 GeV leads to a sizable 
decrease of the decay rates. Therefore, forthcoming data on the absolute 
rates of the isospin-violating decays could shed light on the nature of 
these states: either compact tetraquark states with a scale parameter 
of the order of 2 GeV or more extended objects, which could be viewed 
as hadronic molecules.

\begin{table}
\begin{center}
\caption{Decay rate $\Gamma(B_{s0}^{\ast 0} \to B_s^0 + \pi^0)$ in keV \\ 
at $M_{B_{s0}^{\ast 0}} = 5.725$ GeV.}
\def\arraystretch{.8}
\begin{tabular}{|c|c|c|c|c|}
\hline
& \multicolumn{4}{c|}{$\Gamma(B_{s0}^{\ast 0} \to B_s^0 + \pi^0)$} \\
\cline{2-5}
\hline 
$\Lambda_{B_{s0}^{\ast}}$ [GeV] & I & II & III & Full \\
\hline 
1.00  & 17.3  & 0.12  & 3.7  &  41.3 \\ 
1.25  &  8.7  & 0.06  & 1.9  &  20.7 \\ 
1.50  &  4.4  & 0.03  & 0.9  &  10.4 \\ 
1.75  &  2.3  & 0.01  & 0.5  &   5.4 \\ 
2.00  &  1.2  & 0.01  & 0.3  &   2.9 \\ 
\hline
\end{tabular}
\label{tab:Xbsres1}
\end{center}

\begin{center}
\caption{Decay rate $\Gamma(B_{s0}^{\ast 0} \to B_s^0 + \pi^0)$ in keV \\
at $M_{B_{s0}^{\ast 0}} =6.1$ GeV.}
\def\arraystretch{.8}
\begin{tabular}{|c|c|c|c|c|}
\hline
& \multicolumn{4}{c|}{$\Gamma(B_{s0}^{\ast 0} \to B_s^0 + \pi^0)$} \\
\cline{2-5}
\hline 
$\Lambda_{B_{s0}^{\ast}}$ [GeV] & I & II & III & Full \\
\hline 
1.00  & 19.9  & 0.21  & 4.3  &  48.8 \\ 
1.25  & 12.2  & 0.09  & 2.6  &  29.3 \\ 
1.50  &  6.9  & 0.04  & 1.5  &  16.5 \\ 
1.75  &  3.9  & 0.02  & 0.8  &   9.2 \\ 
2.00  &  2.2  & 0.01  & 0.5  &   5.2 \\ 
\hline
\end{tabular}
\label{tab:Xbsres2}
\end{center}

\begin{center}
\caption{Decay rate $\Gamma(B_{s0}^{\ast 0} \to B_s^0 + \pi^0)$ in keV \\
at $M_{B_{s0}^{\ast 0}} =6.3$ GeV.}
\def\arraystretch{.8}
\begin{tabular}{|c|c|c|c|c|}
\hline
& \multicolumn{4}{c|}{$\Gamma(B_{s0}^{\ast 0} \to B_s^0 + \pi^0)$} \\
\cline{2-5}
\hline 
$\Lambda_{B_{s0}^{\ast}}$ [GeV] & I & II & III & Full \\
\hline 
1.00  & 16.4  & 0.22  & 3.5  &  40.9 \\ 
1.25  & 11.6  & 0.10  & 2.5  &  28.0 \\ 
1.50  &  7.2  & 0.05  & 1.5  &  17.1 \\ 
1.75  &  4.3  & 0.02  & 0.9  &  10.1 \\ 
2.00  &  2.5  & 0.01  & 0.5  &   5.9 \\ 
\hline
\end{tabular}
\label{tab:Xres3}
\end{center}

\begin{center}
\caption{Decay rate $\Gamma(D_{s0}^{\ast +} \to D_s^+ + \pi^0)$ in keV \\
at $M_{D_{s0}^{\ast +}} = 2.317$ GeV.}
\def\arraystretch{.8}
\begin{tabular}{|c|c|c|c|c|}
\hline
& \multicolumn{4}{c|}{$\Gamma(D_{s0}^{\ast +} \to D_s^+ + \pi^0)$} \\
\cline{2-5}
$\Lambda_{D_{s0}^{\ast}}$ [GeV] & I & II & III & Full \\
\hline 
1.00  & 20.5 & 0.15  & 4.4 & 49.1 \\ 
1.25  & 10.5 & 0.08  & 2.2 & 25.1 \\ 
1.50  &  5.4 & 0.04  & 1.1 & 12.8 \\ 
1.75  &  2.8 & 0.02  & 0.6 &  6.7 \\ 
2.00  &  1.5 & 0.01  & 0.3 &  3.7 \\ 
\hline
\end{tabular}
\label{tab:Xcs0}
\end{center}
\end{table}

\begin{acknowledgments}

This work was supported
by the German Bundesministerium f\"ur Bildung und Forschung (BMBF)
under Project No. 05P2015 - ALICE at High Rate (BMBF-FSP 202):
``Jet- and fragmentation processes at ALICE and the parton structure 
of nuclei and structure of heavy hadrons'', 
by Tomsk State University Competitiveness 
Improvement Program and the Russian Federation program ``Nauka'' 
(Contract No. 0.1526.2015, 3854). 
M.A.I.\ acknowledges the support from PRISMA cluster of excellence (Mainz Uni.). 
M.A.I. and J.G.K. thank the Heisenberg-Landau Grant for
support.  

\end{acknowledgments}


\vspace*{-.3cm}

\begin{thebibliography}{99}

\bibitem{Faessler:2007gv} 
  A.~Faessler, T.~Gutsche, V.~E.~Lyubovitskij, and Y.~L.~Ma,
  Phys.\ Rev.\ D {\bf 76}, 014005 (2007).

\bibitem{Faessler:2007us}
  A.~Faessler, T.~Gutsche, V.~E.~Lyubovitskij, and Y.~L.~Ma,
  Phys.\ Rev.\ D {\bf 76}, 114008 (2007).  

\bibitem{Faessler:2008vc}
  A.~Faessler, T.~Gutsche, V.~E.~Lyubovitskij, and Y.~L.~Ma,
  Phys.\ Rev.\ D {\bf 77}, 114013 (2008). 

\bibitem{Agashe:2014kda} 
  K.~A.~Olive {\it et al.}  (Particle Data Group Collaboration),
  Chin.\ Phys.\ C {\bf 38}, 090001 (2014).  

\bibitem{Choi:2015lpc} 
  S.-K.~Choi {\it et al.} (Belle Collaboration),
  Phys.\ Rev.\ D {\bf 91}, 092011 (2015); 
  Phys.\ Rev.\ D {\bf 92}, 039905 (2015).  

\bibitem{Dubnicka:2010kz} 
  S.~Dubnicka, A.~Z.~Dubnickova, M.~A.~Ivanov, and J.~G.~K\"orner,
  Phys.\ Rev.\ D {\bf 81}, 114007 (2010);  
  S.~Dubnicka, A.~Z.~Dubnickova, M.~A.~Ivanov, J.~G.~K\"orner,  
  and G.~G.~Saidullaeva,
  AIP Conf.\ Proc.\  {\bf 1343}, 385 (2011);   
  S.~Dubnicka, A.~Z.~Dubnickova, M.~A.~Ivanov, J.~G.~K\"orner, 
  P.~Santorelli, and G.~G.~Saidullaeva,
  Phys.\ Rev.\ D {\bf 84}, 014006 (2011).  

\bibitem{Ivanov:1996pz}
  M.~A.~Ivanov, M.~P.~Locher, and V.~E.~Lyubovitskij,
  Few-Body Syst.\  {\bf 21}, 131 (1996); 
  M.~A.~Ivanov, V.~E.~Lyubovitskij, J.~G.~K\"orner, and P.~Kroll,
  Phys.\ Rev.\ D {\bf 56}, 348 (1997);  
  I.~V.~Anikin, M.~A.~Ivanov, N.~B.~Kulimanova, and V.~E.~Lyubovitskij,
  Z.\ Phys.\ C {\bf 65}, 681 (1995);
  M.~A.~Ivanov, J.~G.~K\"orner, V.~E.~Lyubovitskij, and A.~G.~Rusetsky,
  Phys.\ Rev.\ D {\bf 60}, 094002 (1999); 
  T.~Branz, A.~Faessler, T.~Gutsche, M.~A.~Ivanov, J.~G.~K\"orner, and V.~E.~Lyubovitskij,
  Phys.\ Rev.\ D {\bf 81}, 034010 (2010); 
T.~Gutsche, M.~A.~Ivanov, J.~G.~K\"orner, V.~E.~Lyubovitskij, 
and P.~Santorelli,
    Phys.\ Rev.\ D {\bf 87}, 074031 (2013). 

\bibitem{Nielsen:2009uh} 
  M.~Nielsen, F.~S.~Navarra, and S.~H.~Lee,
  Phys.\ Rep.\  {\bf 497}, 41 (2010). 


\bibitem{Ivanov:1999pz} 
  M.~A.~Ivanov, J.~G.~Korner, V.~E.~Lyubovitskij, M.~A.~Pisarev,  
  and A.~G.~Rusetsky,
  Phys.\ Rev.\ D {\bf 61}, 114010 (2000). 

 \bibitem{Gasser:1984gg}
  J.~Gasser and H.~Leutwyler,
  Nucl.\ Phys.\  B{\bf 250}, 465 (1985).

\bibitem{Goerke:2016hxf} 
T.~Gutsche, F.~Goerke, M.~A.~Ivanov, J.~G.~K\"orner, 
V.~E.~Lyubovitskij, and P.~Santorelli, 
arXiv:1608.04656 [Phys. Rev. D (to be published)]. 

\end{thebibliography}
\end{document}